\documentclass[aps,pra,pifont,citesort,twocolumn]{revtex4}
\usepackage{lineno}
\usepackage{color,soul}
\usepackage{setspace}
\usepackage[utf8]{inputenc}
\usepackage{graphicx,amsmath,amssymb}
\usepackage{dcolumn}
\usepackage{bm}
\usepackage{times}

\newcommand{\bk}[1]{\left ( #1\right )}

\newcommand{\eqn}[1]{\begin{eqnarray} \newline #1 \end{eqnarray}}
\newcommand{\ee}{&=&}

\newcommand{\hs}{\hspace{0.2cm}}

\newcommand{\bra}[1]{\left \langle#1 \right |}
\newcommand{\ket}[1]{\left |#1\right \rangle}

\newcommand{\EV}[1]{\left < #1 \right >}

\newcommand{\nn}{\nonumber}

\definecolor{darkgreen}{rgb}{0.0, 0.42, 0.24}

\newcommand{\grn}{\color{darkgreen}}
\usepackage{epstopdf} 
\usepackage{epsfig}
\begin{document}

\newcommand{\hmw}[1] {\textcolor{darkgreen}{\small \bf [[#1]]}}
\newcommand{\rep}[2]{\st{#1}{\grn #2}}
\newcommand{\nw}[1] {\textcolor{cyan}{\small \bf [[#1]]}}
\newcommand{\sw}[1] {\textcolor{red}{\small \bf [[#1]]}}

\title{Observation of genuine one-way Einstein-Podolsky-Rosen steering }
\author{Sabine Wollmann$^{1}$}
\author{Nathan Walk$^{1,2}$}
\author{Adam J Bennet$^{1}$}
\author{Howard M Wiseman$^{1}$}
\author{Geoff J Pryde$^{1}$}

\email{g.pryde@griffith.edu.au}

\affiliation{${}^{1}$Centre for Quantum Computation and Communication Technology and Centre for Quantum Dynamics, Griffith University, Brisbane, Queensland 4111, Australia\\${}^{2}$Department of Computer Science, University of Oxford, Oxford OX1 3QD, United Kingdom}

\date{November 4, 2015}

\begin{abstract}
Within the hierarchy of inseparable quantum correlations, Einstein-Podolsky-Rosen steering is distinguished from both entanglement and Bell nonlocality by its asymmetry--- there exist conditions where the steering phenomenon changes from being observable to not observable, simply by exchanging the role of the two measuring parties. Whilst this \textit{one-way steering} feature has been previously demonstrated for the restricted class of Gaussian measurements, for the general case of positive-operator-valued measures even its theoretical existence has only recently been settled. Here, we prove, and then experimentally observe, the one-way steerability of an experimentally practical class of entangled states in this general setting. As well as its foundational significance, the demonstration of fundamentally asymmetric nonlocality also has practical implications for the distribution of the trust in quantum communication networks.

\end{abstract}
\maketitle

Einstein-Podolosky-Rosen steering (or quantum steering) is a nonlocal effect that is distinct from other nonclassical correlations such as Bell nonlocality~\cite{Bell:1964p456} and quantum nonseparability. This distinction manifests as a heirarchy, with each effect witnessed by violation of a corresponding inequality that bounds measurement correlations~\cite{Wiseman:2007p2187}. In this heirarchy, moving from Bell nonlocality to quantum steering to nonseparability requires increasing the number of parties and apparatuses that must be trusted, but the corresponding protocols have been demonstrated to be progressively more robust to noise~\cite{Saunders:2010we,Bennet:2012ch} for projective measurements.
 
 Another logical distinction between steering and the other protocols is immediately apparent from the definitions. For both entanglement and Bell nonlocality, two observers, Alice and Bob, must either jointly share such correlations or not. Steering, however, may be formulated as a quantum information task whereby either Alice or Bob (but not both) is untrusted, but can nonetheless prove the existence of shared entanglement~\cite{Wiseman:2007p2187}. Thus for the asymmetric case of steering, one may ask separately whether Alice can steer Bob, i.e.\ whether or not she can use her measurements to steer his measurement outcomes enough to violate a steering inequality~\cite{Wiseman:2007p2187}, and whether Bob can steer Alice. Finding entangled states for which the two questions have opposing answers, i.e. one-way steerable states, is a highly challenging task. Although a plethora of inequalities exist to demonstrate steering in one direction, to prove no such demonstration exists in the other involves an implicit optimisation over all possible measurement strategies. Thus, the existence of one-way steering considering the general case of positive-operator-valued-measures (POVMs) has only very recently been established in theory~\cite{Quintino:2015uh}, and never observed experimentally. In this Letter we prove the one-way steerability of a readily-accessible class of states which we use to carry out the first observation of truly asymmetric nonlocality.

The existence of asymmetric nonlocal correlations has been observed only under a restricted class of states and measurements---namely, for Gaussian measurements on Gaussian states. Although practical and widely utilised, these Nevertheless, Gaussian resources are provably insufficient for several quantum information applications, including entanglement distillation~\cite{Eisert:2002p466,Fiurasek:2002p467} and universal quantum computation~\cite{Bartlett:2002vx}. Gaussian resources have featured extensively in the development of EPR-steering. Criteria for testing stochastic analogues of the EPR paradox, where position and momentum measurements are made on Gaussian continuous variable states, have long since been proposed~\cite{Reid:1989vm} and experimentally investigated~\cite{Ou:1992p8247}. In this context, it was demonstrated that an asymmetry between the observed violation for Alice and Bob was possible~\cite{Wagner:2008p5277}. With the EPR paradox formalised and generalised as EPR-steering~\cite{Wiseman:2007p2187} (with corresponding steering inequalities), the theoretical existence of one-way steering for Gaussian states and measurements was clear, and was eventually conclusively observed~\cite{Handchen:2012p8278}. 
However, Gaussian measurements are insufficient to capture the full nonlocality of Gaussian states~\cite{Einstein:1935p402,Gisin:1991bu,Gisin:1992wa}. In fact, it is possible to find explicit examples of Gaussian states which are one-way steerable with Gaussian measurements, but two-way steerable when using certain well-chosen non-Gaussian measurements~\cite{Chen:2002jz,supp} (see Appendix A). That is, the presence of 
one-way steerability under a resctricted class of measurements does not imply one-way steerability of the state itself. Do there exist states that are only one-way steerable regardless of the measurements chosen, i.e. genuinely one-way steerable states?

The answer is yes. Following the discovery of example states that were one-way steerable under arbitrary projective measurements~\cite{Bowles:2014cl} and arbitrary finite-setting POVMS~\cite{Skrzypczyk:2014tw}, the theoretical existence of genuine one-way steering was finally settled by Quintino {\em et al.} ~\cite{Quintino:2015uh}. There, an ingenious theorem was used to extend the results of Ref.~\cite{Bowles:2014cl} to infinite-setting POVMs. While conceptually satisfying, these examples belong to a rather exotic family of states and only demonstrate the effect over an extremely small parameter range, making them unsuitable for experimental observation.

Fortunately, a more practical example of one-way steering for an infinite number of arbitrary projective measurements has been independently shown~\cite{Evans:2014bf}. Similarly to Ref.~\cite{Skrzypczyk:2014tw}, it involves distribution through a loss channel, but the authors instead consider the mixture of a singlet state with symmetric noise, i.e. the family of Werner states~\cite{Werner:1989wl}. 

By applying the theorem of Ref.~\cite{Quintino:2015uh}, we are able to construct a family of states that can be steered in one direction with a finite number of Pauli measurements, but, crucially, cannot be steered in the other direction, even for the case of POVMs and infinite settings. Additionally, the experimental tractability of Ref.~\cite{Evans:2014bf} is retained as the new example state also corresponds to a Werner state that has been subjected to a lossy channel. 
 
The demonstration of fundamentally asymmetric nonlocality is of foundational significance in itself. One corollary is that it can be seen as the POVM extension of the results of Ref.~\cite{Saunders:2010we}, namely the observation of steering with Bell local states. This follows because Bell nonlocality can be demonstrated only if bidirectional steering is possible. More practically, as many applications have been found where entanglement~\cite{Horodecki:2009p4263}, Bell nonlocality~\cite{Brunner:2014vw} and steering~\cite{Piani:2015gk,Branciard:2012p8266} respectively play a crucial role, comprehensively resolving these questions provides the ultimate answers as to which entangled states can be seen as resources for which protocols. For example, for the practically relevant case of qubits distributed through lossy, dephasing networks, from the results reported here, one can immediately draw some conclusions about which scenarios  could possibly allow for device-independent quantum key distribution~\cite{Acin:2007p384,Masanes:2011p1898} as opposed to the one-sided device-independent version~\cite{Branciard:2012p8266}.

{\it One-way steering with POVMs.--} Consider two observers, Alice and Bob, performing local measurements on a shared quantum state $\rho$. Alice and Bob have classical strings $k$ and $j$ respectively which label and record the measurements they choose to perform. We will write these measurements as $\{M_{a|k}\}$ and $\{M_{b|j}\}$ where $M_{a|k}$ and $M_{b|j}$ correspond to outcomes $A$ and $B$. For simplicity, the $M_{a|k}$ are often taken to be rank-one projectors, but in general are described by POVM elements that are positive semi-definite ($M_{a|k}\geq 0$) and conserve probability ($\sum_a M_{a|k} = {\bf I}$). We say that the state is steerable if the observed correlations violate any appropriate steering inequality, which is derived from the measurements implemented by the trusted party.

\begin{figure}[htbp]
\includegraphics[width=0.47\textwidth]{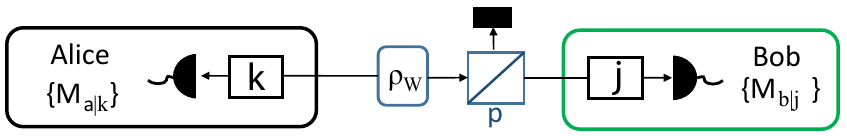}
\caption{Creation of a one-way steerable state (see text for details). One half of a Werner state $\rho_{W}$ is sent directly to Alice, whose measurements are described by $\{M_{a|k}\}$, whilst the other is transmitted to Bob through a loss channel, which replaces a qubit with the vacuum state and is parameterised by probability $p$. Bob's measurements are described by $\{M_{b|j}\}$. For differing values of $p$ the final state is unsteerable by Bob for arbitrary projective measurements or arbitrary POVMs. For the same range of $p$ values, Alice can explicitly demonstrate steering via a finite number of Pauli measurements on both sides. She does this by steering Bob's measurement outcomes so that their shared correlations exceed the upper bound $C_{n}$ allowed in an optimal local hidden state model.}
\label{schem}
\end{figure}

The goal of a one-way steering experiment is to show: (i) that there is no choice of measurements on $\rho_{AB}$ that will allow, say, Bob to demonstrate steering of Alice's state (see Fig.~\ref{schem}); (ii) there exists a specific choice of measurements, $\{M_{a|k}\}$ and $\{M_{b|j}\}$, on the state $\rho_{AB}$ whose output correlations allow Alice to demonstrate steering of Bob's state. 

The steering scenario considered in Ref.~\cite{Evans:2014bf} was for the distribution through a lossy channel of the Werner states defined  as $\rho_{W}=\mu \ket{\psi_s}\bra{\psi_s} + (1-\mu)/4 \hspace{1mm} {\bf I}_4$, 
where $\mu \in [0,1]$, ${\bf I}_4$ is the $4 \times 4$ identity matrix and
$\ket{\psi_s} = \bk{\ket{01} - \ket{10}}/\sqrt{2}$ (ref.~\cite{Evans:2014bf}). These states have been extensively studied in the context of loss tolerant steering inequalities~\cite{Wiseman:2007p2187,Cavalcanti:2009p8246,Evans:2013ce,Evans:2014bf}. These inequalities have been used to definitively demonstrate steering without any detection loophole with a finite number of Pauli measurements~\cite{Bennet:2012ch,Smith:2011cc,Wittmann:2012dg}.

A lossy channel is one that replaces a qubit with the vacuum state $\ket{v}$ with probability $p$ and can be represented by the map $\rho \rightarrow (1-p)\rho + p \ket{v}\bra{v}$.
Specifically, for a Werner state where one subsystem is distributed through a lossy channel to Bob we have the qubit-qutrit state
\eqn{\rho_W \rightarrow \rho_L = (1-p)\rho_W + p \frac{{\bf I}_A}{2}\otimes \ket{v}\bra{v},\label{evans}}
where ${\bf I}_A$ is the identity on Alice's qubit subspace and $\ket{v}$ is a vacuum state orthogonal to Bob's qubit subspace.

Evans and Wiseman observed (endnote 13 of Ref.~\cite{Evans:2014bf}) that the steering inequalities in Ref.~\cite{Bennet:2012ch} demonstrate that, for $\mu \in [1/2,1]$ and if 
\eqn{p>2\mu -1\label{p1},}
 then steering by Bob is impossible even for an infinite number of projective measurements. On the other hand, Alice may steer Bob for any $p$ by simply considering her qubit subspace. Thus Eq.~(\ref{evans}) gives an example of a one-way steerable state if Alice and Bob are restricted to projective measurements.

To make the extension to POVMs we make use of the following result due to Quintino {\em et al.}~\cite{Quintino:2015uh}. If a state, $\tau_{AB}$, is one-way steerable for arbitrary projective measurements then the state \eqn{\rho_{AB} = \frac{1}{d+1}\bk{\tau_{AB} + \pi_\perp\otimes\tau_B} \label{1wayeq}}
is one-way steerable for arbitrary POVMs. Here, $\tau_B = \mathrm{tr}_A\bk{\tau_{AB}}$ and $\pi_\perp$ is a projection operator onto a subspace orthogonal to $\tau_A = \mathrm{tr}_B\bk{\tau_{AB}}$. If we apply this result, setting $\tau_{AB} = \rho_L$ and $\rho_{L_{A(B)}} = \mathrm{tr}_{A(B)}(\rho_L)$, we arrive at the state 
\eqn{\rho_{AB} = \frac{1}{3}\rho_L +   \frac{2}{3}\rho_{L_A}\otimes\pi_\perp,\label{1wayrho}}
which is only one-way steerable, even for arbitrary POVMs. Crucially the orthogonal projection can simply be regarded as transmission through another lossy channel mixing in another vacuum. However, we can effectively combine these into a single loss channel, leading to a final state,
\eqn{\rho_{AB} \ee  \frac{1-p}{3} \rho_W+ \frac{p+2}{3} \frac{{\bf I}_A}{2}\otimes\ket{v}\bra{v}, \label{povmstate}} 
where $\ket{v}$ is the vacuum state. Substituting in Eq.~\ref{p1} we can deduce the relationship between $p$ and $\mu$ for general one-way steering to be
\eqn{p > \frac{2\mu+1}{3}.\label{povm}}

Whilst the above state is provably not steerable by Bob for arbitrary POVMs, we still require an explicit measurement strategy for Alice to demonstrate steering. Due to the erasure channel and inefficient experimental hardware, both Alice and Bob will frequently fail to observe a detection event. In this scenario, Bob is trusted and so he may project into the qubit subspace, making the erasure channel and the efficiency of his detectors irrelevant. Alice, however, must have sufficiently efficient detection to avoid needing to make a fair sampling assumption. \color{black} This is because a dishonest Alice could exploit such an assumption to fake steerable correlations on a subset where she reports an outcome. Demonstrating steering using Werner states in this scenario has been studied, and extremely loss tolerant strategies based upon a finite number of well chosen Pauli measurement settings have been identified and demonstrated~\cite{Bennet:2012ch}. In each round of the demonstration, Bob randomly announces his choice of measurement setting $\hat{\sigma}_{k}^{B}$for $k\in\left\{ {1, ..., n}\right\}$. Alice announces an outcome, $a_{k}\in\left\{{-1,1}\right\}$, which could, in principle, be from her own Pauli measurement or a cheating strategy. After all rounds are complete, Bob computes his steering parameter based upon the correlations and checks whether the inequality
\eqn{S_n = \frac{1}{n} \sum_{k=1}^n  \EV{a_k \hat{\sigma}_{k}^{B}}\leq C_n(\eta_A)}
is violated, where $\eta_A$ is the proportion of rounds where Alice reports an outcome and the bound $C_n(\eta_A)$ is derived from the optimal cheating strategy~\cite{Bennet:2012ch} for that efficiency and choice of measurement settings.

\begin{figure}[htbp]
\includegraphics[width=0.48\textwidth]{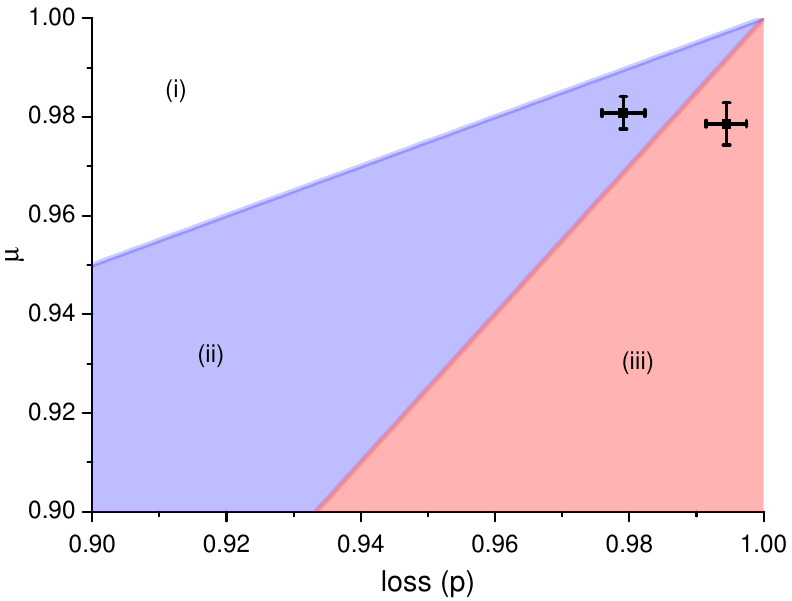}
\caption{The steering regimes are parametrised by the $\mu$-value of the Werner state $\rho_{W}(\mu)$ and the loss $p$. A tunable loss allows the state to be shifted from a regime where it is two-way steerable (i), to a regime where it is one-way steerable if the parties have access to arbitrary projective measurements (ii) and finally a regime where it is one-way steerable even if the parties have access to arbitrary POVMs (iii). The data points with their standard deviations in (ii) and (iii) correspond to the red and blue data points in Fig.~\ref{steering}.}\label{mu}
\end{figure}

By simultaneously verifying that our state (\ref{povmstate}) is steerable by Alice whilst remaining unsteerable by Bob, one can demonstrate one-way steering for both projective and general measurements. The regions for one-way steering for the two scenarios are shown in Fig.~\ref{mu}.

\begin{figure}[htbp]
\includegraphics[width=0.47\textwidth]{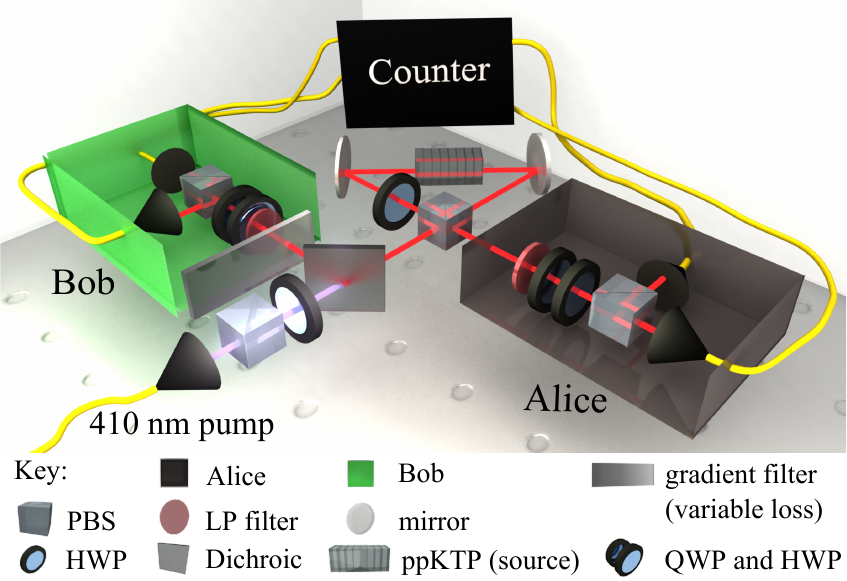}
\caption{In the experimental scheme, Alice and Bob are represented by black and green boxes respectively. Both are in control of their line and their detectors. The party which is steering is additionally in control of the source.
Entangled photon pairs at 820 nm were produced via SPDC in a Sagnac interferometer.
Different measurement settings are realized by rotating half- and quarter-wave plates (HWP and QWP) relative to the polarizing beam splitters. A gradient neutral density (ND) filter is mounted in front of Bob's line to control the fraction of photon qubits passing through. Long pass (LP) filters remove 410 nm pump photons co-propagating with the 820 nm photons before the latter are coupled into single-mode fibres and detected by single photon counting modules and counting electronics.}
\label{setup}
\end{figure}

{\it Experimental details and results.--}
We realised an experimental demonstration of our theory results in a two-photon experiment, where a detection-loophole-free steering violation in one direction was observed to 6 standard deviations. Furthermore, tomographic reconstruction verified the creation of a two-qubit state of the type of Eq.~(\ref{povmstate}), and crucially adhering to the condition expressed in Eq.~(\ref{povm}), which we showed theoretically to be provably unsteerable in the reverse direction even for general POVMs. 
First, polarization Bell states were generated from a high-heralding-efficiency spontaneous parametric down-conversion (SPDC) source, which  enabled us to close the detection loophole (Fig.~\ref{setup}). A fiber-coupled continuous-wave diode laser with $\lambda=410$~nm and output power (after fiber) of 2.5 mW pumped a 1 cm long periodically poled potassium titanyl phosphate (ppKTP) crystal which was mounted in a polarization Sagnac ring interferometer~\cite{Kim2006, Fed2007}.
 At both outputs, we used high-transmission long pass filters to reduce background light, but allowing the SPDC photons at $820$~nm to pass with high efficiency. Furthermore Bob's arm contained a variable neutral density (ND) filter to control the fraction of arriving photons, thereby implementing the lossy channel of Fig.~\ref{schem}(b). Qubit measurements were implemented using wave plates and polarising beam splitters. The photons were coupled into single-mode fibres and detected by Perkin-Elmer single-photon-counting modules (SPCM-AQR-14-FC) with an efficiency of about 50$\%$ at 820 nm. 

To test the quality of the generated entangled photon pair, its joint polarization state was reconstructed via quantum state tomography~\cite{Whi2007} and its fidelity with the closest Werner state, and corresponding parameter $\mu$, were determined (see appendix B). 
We measured the steering parameter $S_{n}$ in our experiment by rotating the HWPs and QWPs for the set of $n$ measurements~\cite{Saunders:2010we,Bennet:2012ch}. We calculated the error for $S_{n}$ as $\Delta S_{n}=\sqrt{\Delta S_{n}\mathrm{(systematic)}^{2}+\Delta S_{n}\mathrm{(statistical)}^{2}}$~\cite{Bennet:2012ch}. The systematic error contribution occurs due to imperfections in Bob's measurement, which could result in overestimating the steering parameter $S_{n}$, while the statistical error is due to Poissonian statistics in coincidence photon counting. EPR-steering experiments usually require that Bob chooses his setting independently from one measurement to the other. Because we controlled Alice's implementation of honest and dishonest strategies, there was no need to force a time ordering of the events. However a strict time ordering has to be reconsidered in a field deployment~\cite{Bennet:2012ch}.

\begin{figure}[htbp]
\includegraphics[width=0.47\textwidth]{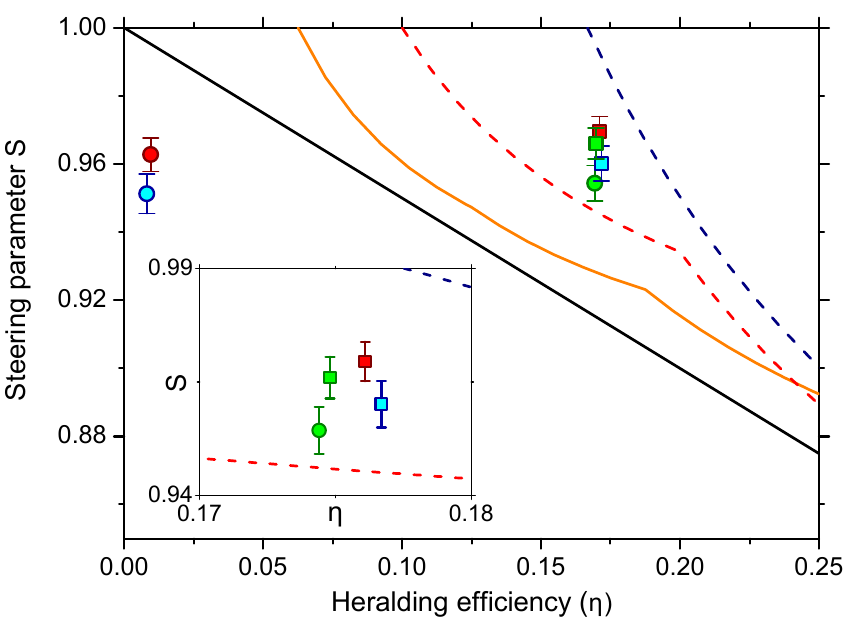}
\caption{Experimental demonstration of one-way EPR-steering. Without any loss Alice (green square) and Bob (green circle) measure a steering parameter $S_{n}$ above the bound for $n=16$ measurements (orange curve; the dashed lines for $n=6$ (blue) and $n=10$ (red) are for information only) and can steer each other by using projective measurements. By adding loss to Bob's arm the Werner state becomes one-way steerable for projective measurements. Alice's steering parameter $S_{16}$ (red square) remains above the bound for $n=16$ measurements, while Bob's (red circle) is below the bound for $n=\infty$ measurements (black), as expected since the state does not allow him to steer Alice by projective measurements. With a significant amount of loss applied, the state becomes such that Bob could not steer Alice even if he could perform arbitrary positive operator-valued measures (POVMs). Both Bob's (blue circle) and Alice's (blue square) measured steering parameters are slightly reduced, but Alice's indicates that she remains able to steer the other party. The horizontal uncertainties are smaller than the data points.} \label{steering}
\end{figure}

First we investigated the case of an EPR-steering task where Alice and Bob could steer each other's state, i.e. a two-way steering task can be completed.
For this, we engineered a heralding efficiency of $\eta_{A}=(16.98\pm0.02)\%$ for Alice and $\eta_{B}=(16.94\pm0.02)\%$ for Bob, which was sufficient to demonstrate steering for $n=16$ measurements, using a dual-platonic-solid arrangement~\cite{Evans:2014bf}.
The generated state had a fidelity of $(99.672\pm0.001)\%$ with the nearest Werner state with $\mu=0.991\pm0.002$. The parameter $\mu$ was engineered to be sufficiently high such that the subsequent highly correlated state allowed EPR-steering at an experimentally accessible heralding efficiency.
We successfully violated the inequality in both steering directions with $S_{16}=0.966\pm0.005$ for Alice and $S_{16}=0.954\pm0.005$ for Bob (Fig.~\ref{steering}). The steering parameters were 8.4 standard deviations for Alice, and 5.1 standard deviations for Bob, above the bound.

Adding a ND gradient filter into Bob's beam shifted the state into a regime where it was one-way steerable for projective measurements. For this, we arranged a state with a fidelity of $(99.6\pm0.1)\%$ with a Werner state of $\mu=0.991\pm0.003$ and applied loss with $p=(87\pm3)\%$ (Fig.~\ref{mu}).
Alice remained able to steer the other party with $S_{16}=0.970\pm0.004$, 7.3 standard deviations above the bound, at $\eta=(17.11\pm0.07)\%$ (Fig.~\ref{steering}). The loss of information in Bob's arm made him unable to steer the other party. We observed a steering parameter of $S_{16}=0.951\pm0.006$. In this case, this S value would not have violated a steering inequality even with an infinite number of measurements.

Finally, we investigated the regime where only one-way steering is possible, even for arbitrary POVMs. We produced, and completely characterised by tomography, a state having fidelity of $(99.1\pm0.3)\%$ with a Werner state of $\mu=0.978\pm0.008$ with an applied loss of $p=(99.5\pm0.3)\%$ (Fig.~\ref{mu}). 
Alice remained able to steer Bob with a steering parameter $S_{16}=0.960\pm0.005$, being 6.6 standard deviations above the bound, at $\eta=(17.17\pm0.04)\%$ (Fig.~\ref{steering}). Bob's steering parameter $S_{16}=0.951\pm0.006$ did not violate the inequality (Fig.~\ref{steering}) and there is no kind of measurement he could choose, even in principle, to be able to steer Alice.                                                                    

{\it Conclusions.--} 

We have conclusively demonstrated the phenomena of one-way steering in the general setting of non-sequential POVMs, providing a clear demonstration of the inequivalence between steering, entanglement and Bell nonlocality. As an immediate consequence, we have identified a class of channels which could never allow fully device-independent QKD with Werner states for any measurement strategy but may still permit one-sided device-independent protocols.

Several natural extensions to this work remain. Of primary interest is the question of whether the bound for one-way steering with POVMs derived here is in fact tight. We conjecture that it is not, based upon comparison with the work of Skrzpczyk {\em et al.}~\cite{Skrzypczyk:2014tw}. Considering their results suggests that the bounds for one-way steering with arbitrary projective measurements may well hold for arbitrary POVMs. This approach may also hold promise for resolving a longstanding open problem in the study of in Werner states~\cite{Werner:2014vg}, namely, for what values of $\mu$ are they steerable or nonlocal?

Another interesting avenue would be to investigate the nature of asymmetric nonlocality in the multipartite setting for general POVMs. One might also consider the most general possible measurement strategies, involving sequences of measurements. Finally, it would illuminating to extend this analysis to higher dimensions, in particular to consider the one-way steerability of the Gaussian continuous variable states of the original EPR argument.

We acknowledge helpful discussions with Michael Hall, Raj Patel, Sergei Slussarenko and Helen Chzranowski. This research was supported by the ARC Centre of Excellence CE110001027. NW acknowledges support from the EPSRC National Quantum Technology Hub in Networked Quantum Information Technologies.

\appendix
\section{Steering Gaussian states with non-Gaussian measurements\label{nongauss}}
Here, we present an explicit example of a two-way steerable state that appears one-way steerable if we restrict to Gaussian measurements. We consider the canonical EPR state, which is usually created by combining squeezed states. It can be written in the number basis as
\eqn{\ket{\Psi_{\mathrm{EPR}} } = \sqrt{1-\chi^2}\sum_{n=0}^\infty \chi^n \ket{n,n},} where $\chi$ parameterises the entanglement and is related to the variance of the squeezed resource states, $V_{\mathrm{sq}}$, via \eqn{\chi = \sqrt{\frac{V_{\mathrm{sq}}-1}{V_{\mathrm{sq}}+1}} .}  
For infinite squeezing $\chi$ goes to 1 and we have a maximally entangled state. For Gaussian quadrature measurements $X_A,P_A$ ($X_B,P_B$) made by Alice (Bob) steering can be demonstrated by violating the Reid criteria on the conditional variances~\cite{Reid:1989vm},
\eqn{V_{X_A|X_B}V_{P_A|P_B}\geq 1}
for Bob to steer Alice and
\eqn{V_{X_B|X_A}V_{P_B|P_A}\geq 1}
for Alice to steer Bob.
If Bob's arm is transmitted through a lossy channel of transmission $T$, then Alice can steer Bob for any $T$ but Bob may only steer Alice for $T>1/2$. However, we consider instead the measurements introduced used by Chen {\em et al.} in the context of Bell tests. These are infinite dimensional analogues of the Pauli operators and are given by
\eqn{S_Z \ee\nn \sum_{n=0}^\infty \ket{2n+1}\bra{2n+1} -\ket{2n}\bra{2n},\\
S_- \ee  \sum_{n=0}^\infty \ket{2n+1} \bra{2n} = (S_+)^\dag \nn,\\
S_{\pm}\ee S_X \pm i S_Y .}
We apply these infinite dimensional but dichotomic (all outcomes are either 1 or -1) measurements to a non-linear steering inequality derived by Jones and Wiseman~\cite{Jones:2011kh}. They consider an equatorial family of measurements,
\eqn{S_{\theta} = \cos(\theta) S_X + \sin(\theta) S_Y}
along with the $S_Z$ measurement. For the case where Alice is trying to steer Bob we define the variable $A_\theta,A_Z$ as Alice's reported measurement outcome when Bob specifies $S_\theta,S_Z$. If we further define $P_+,P_-$ as the probabilities that Alice announces $A_Z=1,A_Z=-1$, and $Z_+,Z_-$ Bob's conditional expectation values, then steering is demonstrated by the violation of the inequality
\eqn{\int_{-\pi}^\pi d\theta \hs \EV{A_\theta S_{\theta}} &\leq&  \frac{2}{\pi} \bk{P_+\sqrt{1-Z_+^2} + P_-\sqrt{1-Z_-^2}}\nn
\\
\label{ngsteer}}
A dishonest Alice could use any strategy to announce variables $A_\theta,A_Z$ whereas an honest Alice will simply measure the corresponding observables on her side. Interestingly, because of the dependence upon Alice's announced values, in an experiment with an honest Alice the value of the bound will itself change as a function of the state.

Steering in the other direction can be shown via the same inequality with the roles of Alice and Bob interchanged. Once again, note that contrary to the conditional variance inequalities, it is the bound on the right-hand side that is different when Bob is trying to steer Alice whereas the expression on the left-hand side is the same in both instances.

\begin{figure}[htbp]
\includegraphics[width=0.47\textwidth]{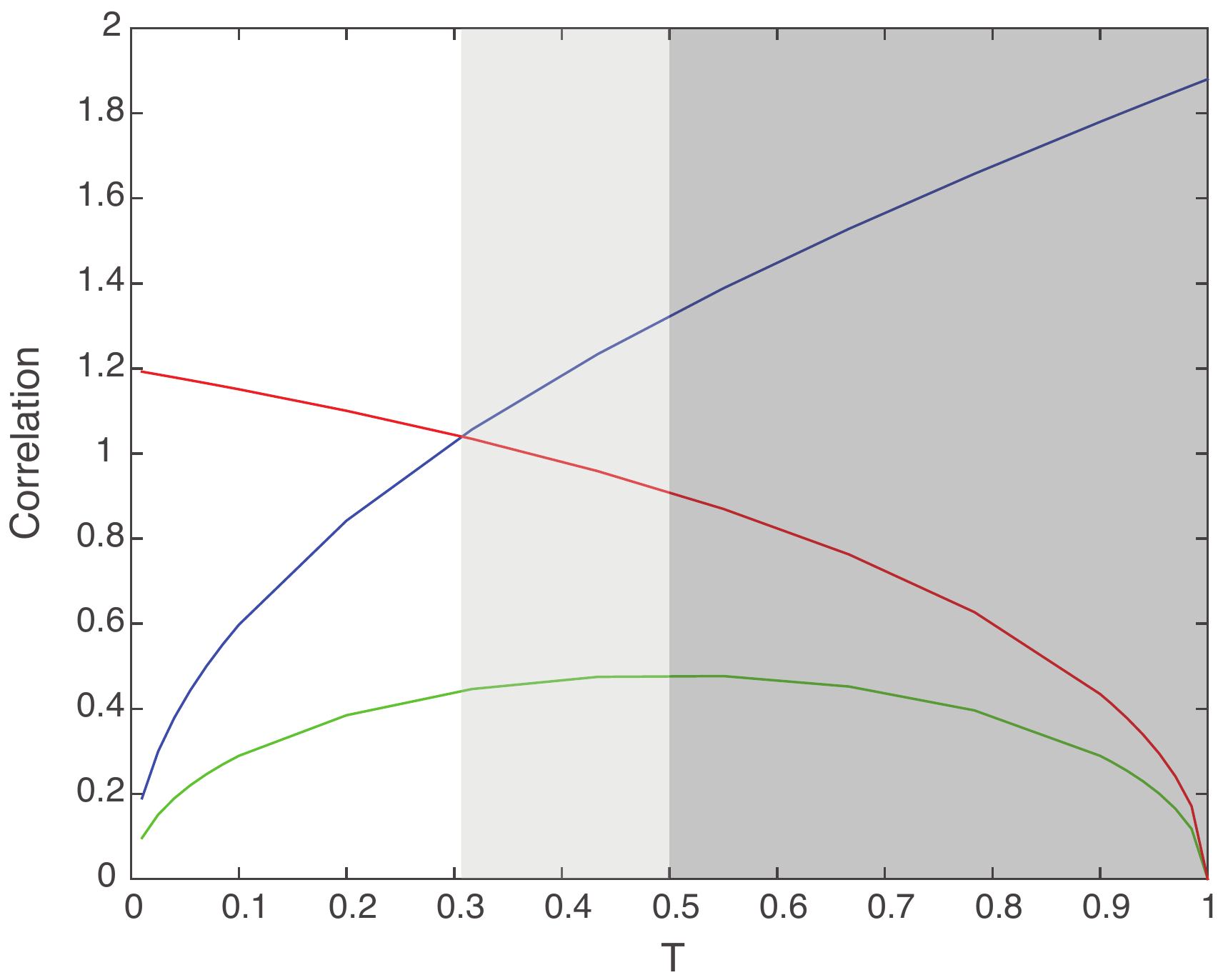}
\caption{Two way steering of a state that is one-way steerable with Gaussian measurements as a function of the transmission to Bob, $T$. Steering is demonstrated when the l.h.s. of Eq.~\ref{ngsteer} (blue curve) is greater the r.h.s (green curve for Bob to steer Alice or red curve for Alice to steer Bob). For initial pure two-mode squeezing of 3dB, the state is demonstrably two-way steerable for $T\gtrsim 0.3$, shown by light grey shading. For comparison, the two-way steering region for Gaussian measurements is shaded in dark grey.}\label{2way}
\end{figure}

Using this expression, we analyse an EPR state with modest 3~dB of squeezing. We find that in this case, the EPR state is in fact two-way steerable for $T\gtrsim 0.3$ (Fig.~\ref{2way}), thereby showing that one-way steering under a Gaussian measurements restriction can sometimes vanish with more general measurements. 

\section{Determination of the experimental Werner parameter $\mu$}
The quantum states that we create are Werner-like, well-described by 
\begin{equation}
{\rho=(\hat{U}\otimes {\bf I})W_{\mu}(\hat{U}\otimes {\bf I})^{\dagger},}
\end{equation}
where $\hat{U}$ is a single-qubit unitary operation and $W_\mu$ is the density matrix of a Werner state. By incorporating a unitary transformation in the measurement settings of qubit 1, we can retrieve a Werner state~\cite{Saunders:2010we}. The search for the optimal unitary operation is based upon the tomographic reconstruction of $\rho$~\cite{Jam2001}. The optimal $\hat{U}$ is determined by searching over all possible unitary operations to numerically maximize the fidelity of $(\hat{U}\otimes {\bf I})\rho(\hat{U}\otimes {\bf I})^{\dagger}$ with the closest Werner state $W_{\mu}$. Our final result is given by the minimum of the cost function 
\begin{equation}
 \textrm{cost}=\min_{\mu,U}\left(1- \mathcal{F}\left(\left(\hat{U}\otimes {\bf I}\right)^{\dagger}\rho\left(\hat{U}\otimes {\bf I}\right), W_{\mu}\right)\right).
\end{equation}
The fidelity for mixed states is defined by 
\begin{equation}
 \mathcal{F}= \textrm{Tr}\left[\sqrt{\sqrt{\rho} \sigma \sqrt{\rho}})\right]^2 
\end{equation}
 where $\rho$ and $\sigma$ are the density matrices of the states being compared.
\bibliographystyle{apsrev}


\end{document}